\documentclass[final,3p,times,twocolumn]{elsarticle}

\usepackage{graphicx}
\usepackage{amssymb}

\usepackage{lineno}

\biboptions{sort&compress}
\usepackage{algorithmic}
\usepackage[ruled,vlined,commentsnumbered]{algorithm2e}
\usepackage{multirow}
\usepackage{array}
\usepackage{mdwmath}
\usepackage{mdwtab}
\usepackage{eqparbox}
\usepackage{url}
\usepackage[cmex10]{amsmath}

\journal{Computers in Biology and Medicine}

\begin{document}

\begin{frontmatter}


\title{A Novel Self-Learning Framework for Bladder Cancer Grading Using Histopathological Images}

\ead{jogarpa7@i3b.upv.es}



\author{Gabriel Garc\'ia\textsuperscript{1}, Anna Esteve\textsuperscript{1,2}, Adri\'an Colomer\textsuperscript{1}, David Ramos\textsuperscript{2} and Valery Naranjo\textsuperscript{1}}

\address{\textsuperscript{1}Instituto de Investigación e Innovación en Bioingeniería, Universitat Politècnica de València, 46022, Valencia, Spain}
\address{\textsuperscript{2} Hospital Universitario y Politécnico La Fe, Avinguda de Fernando Abril Martorell, 106, 46026, Valencia, Spain.}

\begin{abstract}

Recently, bladder cancer has been significantly increased in terms of incidence and mortality. Currently, two subtypes are known based on tumour growth: non-muscle invasive (NMIBC) and muscle-invasive bladder cancer (MIBC). In this work, we focus on the MIBC subtype because it is of the worst prognosis and can spread to adjacent organs. We present a self-learning framework to grade bladder cancer from histological images stained via immunohistochemical techniques. Specifically, we propose a novel Deep Convolutional Embedded Attention Clustering (DCEAC) which allows classifying histological patches into different severity levels of the disease, according to the patterns established in the literature. The proposed DCEAC model follows a two-step fully unsupervised learning methodology to discern between non-tumour, mild and infiltrative patterns from high-resolution samples of $512\times 512$ pixels. Our system outperforms previous clustering-based methods by including a convolutional attention module, which allows refining the features of the latent space before the classification stage. The proposed network exceeds state-of-the-art approaches by $2-3\%$ across different metrics, achieving a final average accuracy of $0.9034$ in a multi-class scenario. Furthermore, the reported class activation maps evidence that our model is able to learn by itself the same patterns that clinicians consider relevant, without incurring prior annotation steps. This fact supposes a breakthrough in muscle-invasive bladder cancer grading which bridges the gap with respect to train the model on labelled data.

\end{abstract}


\end{frontmatter}


\section{Introduction}\label{sec:1_0_Introduction}

Bladder cancer is an uncontrolled proliferation of the urothelial bladder tumour cells that entails the development of the tumour. A significant increase in adult incidence and mortality has been observed during the last years regarding this alteration. In particular, recent studies claim that bladder cancer is the second most common urinary tract cancer and the fifth most incident among men in developed countries \cite{antoni2017, lorenzo2018}.

Nowadays, the diagnostic procedure of bladder cancer includes several time-consuming tests. First, urine cytology is performed to determine the presence of carcinogenic cells \cite{feil2006}. Later, a vesico-prostatic and renal ultrasound is employed to locate the tumour and get an idea about its kind of growing, which provides relevant cues to determine the grade and the prognosis of the patient. When the tumour can not be located in the previous stage, an MRI urogram is carried out to analyze possible local spread \cite{sharma2009}. If there is evidence of bladder cancer, the urologist usually performs a cystoscopy based on the transurethral resection (TUR) technique, which allows extracting a sample of abnormal bladder tissue to determine the kind of tumour growth. After the preparation process, the biopsied tissue is usually stained with hematoxylin and eosin ($H\&E$) to enhance the histological properties of the tissue. Finally, an additional staining process can be adopted to highlight special structures associated with the problem under study. Particularly, the immunohistochemical CK AE1/3 technique was applied on the histological images used in this work to highlight the carcinogenic cells by providing a brown hue when the antigen-antibody binding occurs. 

\begin{figure*}[h]
\begin{center}
    \includegraphics[width=16cm]{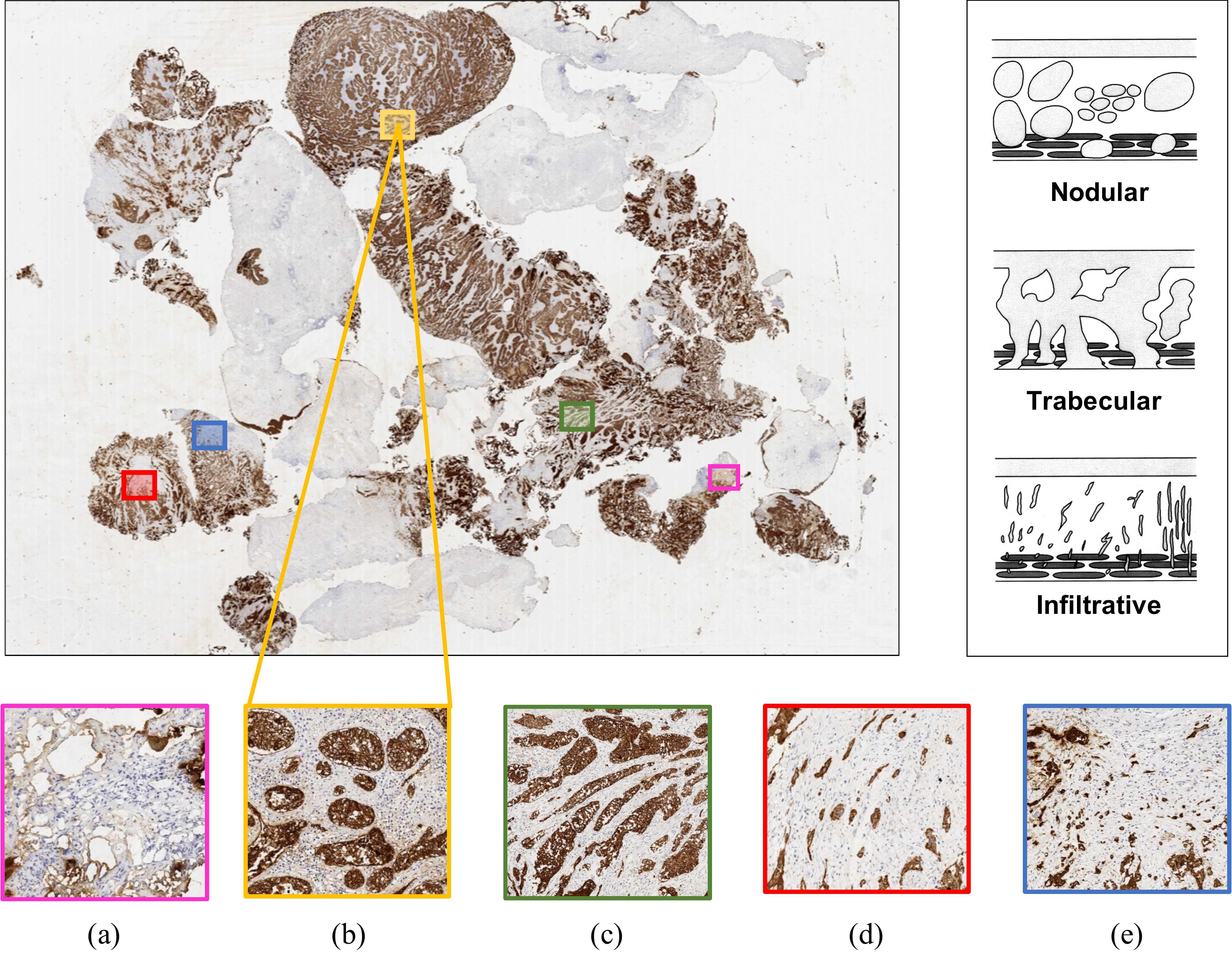}
\end{center}
\caption{Whole Slide Image (WSI) from a patient suffering from muscle-invasive bladder cancer (MIBC) in which different growth patterns are evidenced. (a) Non-tumour pattern. (b) Nodular arrangement (mild pattern). (c) Trabecular arrangement (mild pattern). (d) Tumour cell strands of an infiltrative pattern. (e) Isolated tumour cells denoting an infiltrative pattern.}
\label{fig_bladder_patterns}
\end{figure*}

Note that two kinds of bladder cancer, non-muscle invasive (NMIBC) and muscle-invasive (MIBC), are distinguished depending on its level of invasiveness during the tumour growth within the bladder wall. Currently, 75\% and 25\% of the bladder cancer cases correspond to NMIBC and MIBC, respectively \cite{lorenzo2018}. In this study, we focus on the MIBC category since it leads to the worst prognosis and favours the spread of the tumour to adjacent organs. According to \cite{stenzl2010}, the muscle-invasive bladder cancer (MIBC) does not usually present low-grade cases of malignancy, but just high-grade urothelial carcinomas, which can be categorized as grade 2 or 3 following the classification criteria proposed by the World Health Organization (WHO) \cite{busch2002}. Jimenez et al. \cite{jimenez2000} described three different histological patterns which keep correlation with the patient outcome. Specifically, nodular, trabecular and infiltrative patterns can be found in the histopathological images stained with CK AE1/3, as observed in Figure \ref{fig_bladder_patterns}. The nodular pattern (yellow box) is defined by well-delineated tumour cell nests with a circular shape. Otherwise, the trabecular pattern is characterised by the presence of tumour cells arranged in interconnected bands. Finally, the infiltrative pattern is composed of tumour cell strands (red box) or a small set of isolated cells called \textit{buds} (blue box). The infiltrative pattern, a.k.a. tumour budding, represents the most aggressive scenario and the worst prognosis for the patient \cite{almangush2016, karamitopoulou2013, masuda2012, fukumoto2016}. For that reason, we pigeonholed nodular and trabecular structures in a single specific class (mild pattern) to grade the MIBC severity according to the prognosis of the disease. Also, we considered a non-tumour pattern (pink box) to cover those cases in which the patient does not present signs of tumour evidence. In this way, a multi-class scenario is conducted throughout the paper to grade the bladder cancer into non-tumour (NT), mild (M) and infiltrative (I) patterns.

    \subsection{Related work}\label{subsec:1_1_Related_work}

An accurate bladder cancer diagnosis supposes a very time-consuming task for expert pathologists, whose level of reproducibility is low enough to provide significant differences in the histological-based interpretation \cite{wetteland2019, zhang2019}. For that reason, many studies in the state of the art have proposed artificial-intelligence algorithms to help pathologists in terms of cost-effectiveness and subjectivity ratio. Most of them focused on machine-learning techniques applied on $H\&E$-stained histological images for segmentation \cite{lucas2020, yin2020, dolz2018} and classification \cite{woerl2020, xu2019, harmon2020, wetteland2019, zhang2019, yang2021, ikeda2020, yang2020} problems.

Regarding the segmentation-based studies, Lucas et al. \cite{lucas2020} used the popular U-net architecture to segment normal and malignant cases of bladder images. Then, they used the common VGG16 network as a backbone to extract histological features from patches of $224\times224$ pixels. The resulted features were fused with other clinical data to address a classification stage via bidirectional GRU networks. The proposed algorithm reported accuracy of 0.67 for the 5-year survival prediction. In \cite{yin2020}, the authors carried out an end-to-end approach to discern between MIBC and NMIBC categories from $H\&E$ images. First, they performed a segmentation process to discriminate the tissue from the background of the image. Patches of $700\times700$ pixels were used to perform both manual and automatic feature extraction. The hand-crafted learning was conducted via contextual features such as nuclear size distribution, crack edge, sample ratio, etc., whereas the data-driven learning was addressed via VGG16 and VGG19 architectures. During the classification stage, different machine-learning classifiers such as support vector machine (SVM), logistic regression (LR) or random forest (RF), among others, were used to determine the kind of bladder tissue. The hand-driven approaches showcased an outperforming with respect to the deep-learning models, achieving an accuracy of $91-96\%$ depending on the classifier.

Respecting the classification-intended studies, most of them were focused on $H\&E$-stained histological images, as before. In \cite{woerl2020}, the researchers proposed a multi-class scenario to detect the molecular sub-type in muscle-invasive bladder cancer (MIBC) cases. They applied the ResNet architecture on patches of $512\times512$ pixels, achieving results for the area under the ROC curve (AUC) of 0.89 and 0.87 in terms of micro and macro-average. Otherwise, in \cite{xu2019}, the authors made use of the Xception network as a feature extractor from $H\&E$-stained patches of $256\times256$ pixels. Then, an SVM classifier was implemented to discern between high and low mutational burden reaching values of 0.73 and 0.75 for accuracy and AUC metrics, respectively. Harmon et al. \cite{harmon2020} proposed a classification scenario to detect lymph node metastases from $H\&E$ patches of $100\times100$ pixels. A combination of the ResNet-101 architecture with AdaBoost classifiers reported an AUC of 0.678 at the test time. Another interesting study \cite{wetteland2019} carried out a classification approach to categorize the kind of the tissue into six different classes: urothelium, stroma, damaged, muscle, blood and background. To this end, the authors combined supervised and unsupervised deep-learning techniques on patches of $128\times128$ pixels stained with $H\&E$. Specifically, they trained an autoencoder (AE) from the unlabelled images and used the encoder network to address the classification throughout the features extracted from the labelled samples. They reached multi-class results of 0.936, 0.935 and 0.934\% for precision, recall and F1-score metrics, respectively. One of the more important state-of-the-art studies focused on $H\&E$ histological images of bladder cancer was carried out in \cite{zhang2019}. Zhang et al. collected a large database of WSIs with the aim of discerning between low and high grades of the disease. In particular, they used an autoencoder network to identify possible areas with cancer. Then, the extracted ROIs of dimensions $1024\times1024$ pixels were input to a Convolutional Neural Network (CNN) to classify them into low and high classes. Finally, an average accuracy of $94\%$ was reported by the proposed system, in comparison to the $84,3\%$ reached by the pathologists. The findings from this study reveal that there exists a significant subjectivity level between experts in the diagnosis from histological bladder cancer images, as supported in \cite{wetteland2019}.

It should be noted that, besides the histopathological samples, other imaging modalities are also considered in the literature for bladder cancer analysis, e.g. magnetic resonance imaging (MRI) \cite{dolz2018}, cystoscopy \cite{ikeda2020, yang2020} or computerized tomography (CT) \cite{yang2021}. Particularly, Dolz et al. \cite{dolz2018} applied deep-learning algorithms to detect bladder walls and tumour regions from MRI samples. In \cite{ikeda2020, yang2020}, different deep-learning architectures were implemented to distinguish between healthy and bladder cancer patients using cystoscopy samples. Yang et al. \cite{yang2021} outlined a classification between NMIBC and MIBC categories from CT images. Additionally, although immunohistochemical techniques are widely used in the literature for detecting tumour budding, most of the state-of-the-art works applied them on colorectal cancer images \cite{prall2005, lugli2009, ogawa2009, zlobec2012}. However, a large gap in immunohistochemical-based studies is found in the literature for bladder cancer diagnosis. As far as we know, only the study carried out in \cite{brieu2019} proposed the use of immunofluorescence-stained samples to quantify the tumour budding for MIBC prognosis via machine learning algorithms. Specifically, the authors aimed to establish a relationship between tumour budding and survival evaluated in patients with MIBC. To this end, they carried out learning strategies based on nuclei detection and segmentation of the tumour in stroma regions to count the isolated tumour budding cells. The authors proposed a survival decision function based on random forest classifiers, reporting a hazard ratio of 5.44.

    \subsection{Contribution of this work}\label{subsec:1_2_Contribution}

To the best of the author's knowledge, no previous works have been performed to analyse the severity of bladder cancer using histological images stained with cytokeratin AE1-AE3 immunohistochemistry. Moreover, all the state-of-the-art studies focused on supervised learning methods to find dependencies between the inputs and the predicted class \cite{wetteland2019, zhang2019, xu2019, harmon2020}. Some of them \cite{wetteland2019, zhang2019} also considered the use of unsupervised techniques in early methodological steps to find possible ROIs with cancer, but they need labelling data to build the definitive predictive models. Besides, pattern recognition tasks aimed to grade bladder cancer have not been addressed in previous studies. 

To fill these gaps in the literature, we present in this paper a self-learning framework for bladder cancer growth pattern, which focuses on fully unsupervised learning strategies applied on CK AE3/1-stained WSIs. In particular, we propose a deep convolutional embedded attention clustering (DCEAC) which allows boosting the performance of the classification model without incurring labelled data. In the literature, deep-clustering algorithms have demonstrated a high rate of performance for image classification \cite{guo2017, guo2018, xie2016}, image segmentation \cite{enguehard2019}, speech separation \cite{hershey2016, prasetio2019} or data analysis \cite{rocio2021}, among other tasks. Inspired by \cite{guo2017}, we propose a tailored algorithm capable of competing with the state-of-the-art results achieved by supervised algorithms. As a novelty, we include a convolutional attention module to refine the features embedded in the latent space. Additionally, we are the first that focus on the arrangement of the histological structures contained in the high-resolution patches to classify them into non-tumour (NT), mild (M) and infiltrative (I) patterns, according to the criteria proposed in \cite{jimenez2000}. We also compute a class activation map (CAM) algorithm \cite{zhou2016} to evidence how the proposed network pays attention to those specific structures which match with the clinical patterns associated with the aggressiveness of bladder cancer.

With all of the above, the proposed end-to-end framework supposes a reliable benchmark for making diagnostic suggestions without involving the pathologist experience, which adds significant value to the body of knowledge. In summary, the main contributions of this work are listed below:

\begin{itemize}
\item For the first time, we make use of CK AE3/1-stained images to address the automatic diagnosis of bladder cancer via machine-learning algorithms.

\item We resort to advanced unsupervised deep-learning techniques to address the bladder cancer grading without the need for prior annotation steps. 

\item We propose a novel deep-clustering architecture able to improve the representation space via convolutional attention modules, which derives in a better-unsupervised classification. 

\item We based on high-resolution histological patches to learn specific-bladder cancer patterns and stratify the different severity levels of the disease according to the literature.

\item Heat maps highlighting decisive areas are reported to incorporate an explainable component for the network prediction. This fact provides an interpretability perspective that coincides with the clinicians' criteria.
\end{itemize}

\section{Material}\label{sec:2_0_Material}

A private database composed of 136 whole-slide images (WSIs) stained via immunohistochemistry CK AE1/3 technique was used to accomplish this study. The WSIs, coming from the \textit{Hospital Universitario y Politécnico La Fe (Valencia, Spain)}, were digitized at the highest optical magnification ($40\times$) to leverage the inherent structure of the bladder patterns associated with each grade of the disease. Worth noting that a high image resolution is necessary to achieve an accurate diagnosis of bladder cancer. This is because the class dependencies are evidenced in the high frequency of the image, especially the tumour budding details. 

\begin{figure*}[t]
\begin{center}
    \includegraphics[width=17cm]{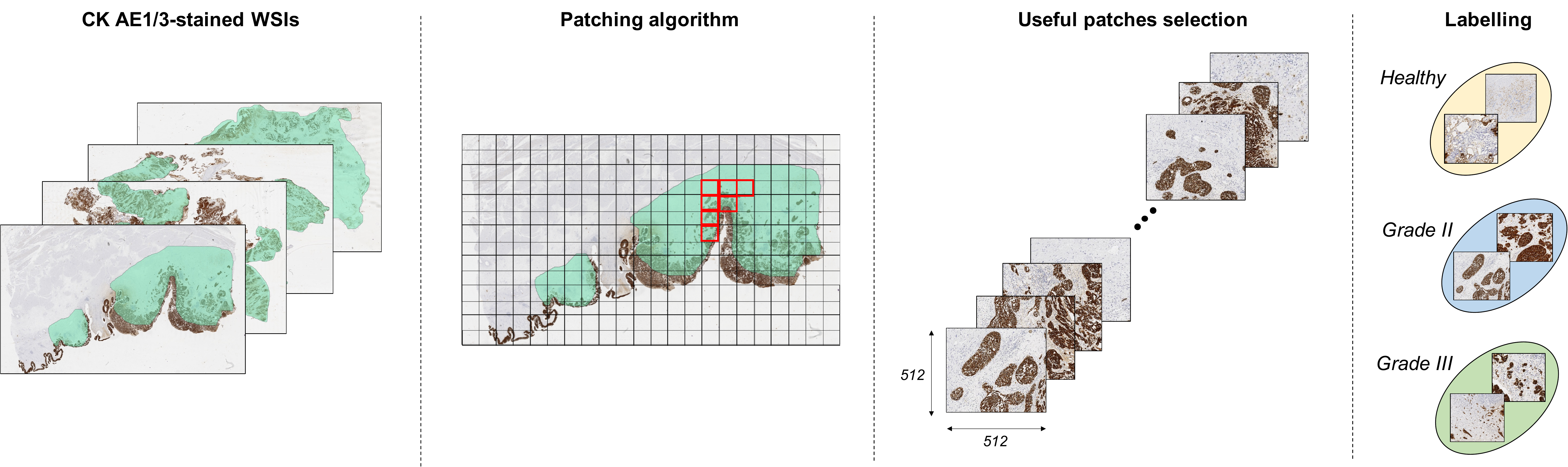}
\end{center}
\caption{Database preparation process. First, a patching algorithm was applied on 136 CK AE1/3-stained WSIs to extract sub-images of $512\times512$ pixels. For validation purposes, the resulting 2995 patches were labelled by an expert as non-tumour (NT), mild (M) or infiltrative (I) pattern to give rise to a multi-class scenario for bladder cancer grading.}
\label{fig_material}
\end{figure*}

In the first step of the database preparation, an expert from the Pathological Anatomy Department carried out a manual segmentation to indicate possible areas of interest. At this point, it is important to highlight that the segmentation was performed in a very rough manner, as observed in Figure \ref{fig_material}, in order to reduce the expert's annotation time as much as possible. From here, a patching algorithm was applied to extract cropped images with an optimal block size in terms of computational efficiency and structural content. Specifically, we extracted patches of dimensions $512\times512$ pixels, according to some of the most recent state-of-the-art studies focused on histopathological images \cite{rho2021, silva2021, woerl2020}. Then, a preprocessing step was addressed to discard the useless regions, corresponding to the background of the WSI, and select those patches containing more than 75\% of annotated tissue. After this, a total of 2995 representative patches composed the unsupervised framework. For validation purposes, an expert manually labelled each patch as non-tumour (NT), mild (M) or infiltrative (I) classes, according to the pattern criteria previously detailed in Section \ref{sec:1_0_Introduction}. The labelling process resulted in 763 non-tumour, 1470 mild and 762 infiltrative cases, as reported in Figure \ref{fig_material}. It is essential to remark that we did not have access to the labelled data during the training phase since we propose a fully unsupervised strategy to achieve a self-learning of the patterns. Labels were just considered at the test time to evaluate the models' performance. 

\section{Methods}\label{sec:03_Methods}

Recently, deep-clustering algorithms have risen to the forefront of unsupervised image-based techniques since they allow enhancing the feature learning while improving the clustering performance in a unified framework \cite{guo2017}. In this paper, we address a fully unsupervised self-learning strategy to cluster a large collection of unlabelled images into $K=3$ groups corresponding to different severity levels of MIBC. Specifically, inspired by \cite{guo2017}, we propose a novel deep convolutional embedded attention clustering (DCEAC) in which the features are updated in an online manner to learn stable representations for the clustering stage. Unlike conventional approaches \cite{xie2016}, the proposed DCEAC algorithm optimizes the latent space by preserving the local structure of data, which helps to stabilize the clustering-learning process without distorting the embedding properties. 

Self-learning methods aim to learn useful representations by leveraging the domain-specific knowledge from the unlabelled data to accomplish downstream tasks. This training procedure is usually faced by solving pretext tasks \cite{gidaris2018}, relational reasoning \cite{patacchiola2020} or contrastive learning \cite{chen2020} approaches. In our bladder cancer scenario, we advocate for a sequential strategy that resorts to image reconstruction as an unsupervised pretext task. Specifically, we conduct a two-step learning methodology in which, first, a convolutional autoencoder (CAE) is trained to incorporate information about the domain properties of the histological patches. In a second phase, a clustering branch is included at the output of the CAE bottleneck to provide the class information from the embedded features, which are online updated by re-training the CAE in a combined network. Below, we detail both learning steps.

\subsection{CAE pre-training} \label{subsec:3_1_CAE_pretraining}

Autoencoder (AE) is one of the most common techniques for data representation, whose aim is to minimize the reconstruction error between inputs $X$ and outputs $R$. AE architectures are composed of two training stages: encoder $f_\phi(\cdot)$ and decoder $g_\theta(\cdot)$, where $\phi$ and $\theta$ are learnable parameters. The encoder network applies a non-linear mapping function to extract a feature space $Z$ from the input samples $X$, so that $f: X \rightarrow Z$. Then, the decoder structure is intended to reconstruct the input data from the embedded representations via $R = g_\theta(Z)$. The learning procedure is carried out by minimizing a reconstruction loss function.

\begin{figure*}[h]
\begin{center}
    \includegraphics[width=14.5cm]{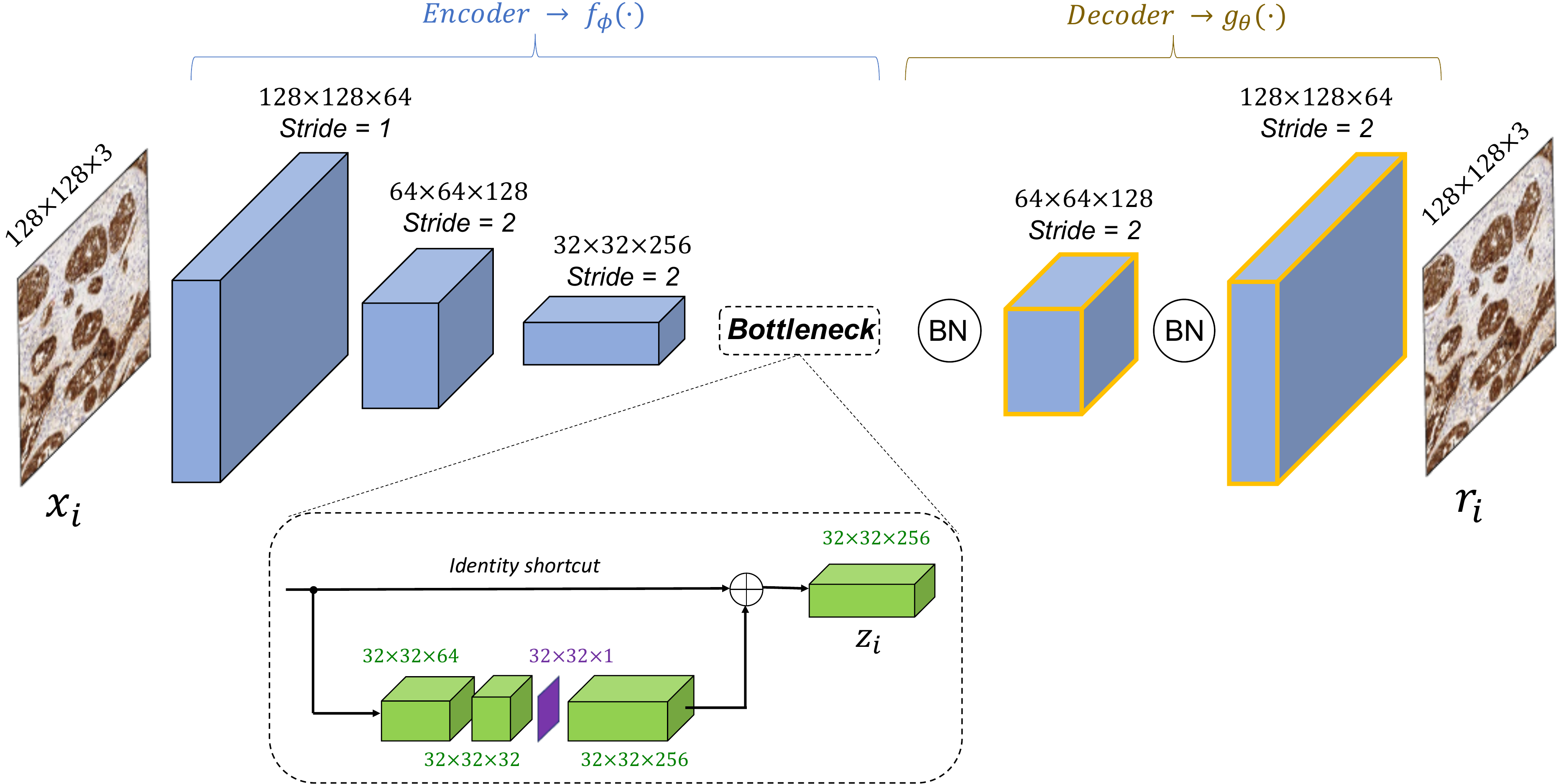}
\end{center}
\caption{Architecture of the proposed CAE used for image reconstruction as a pretext task during the learning process.}
\label{fig_CAE}
\end{figure*}

Notice that AE architectures are usually defined by fully connected layers, intended to reduce the dimensionality of the feature space \cite{rocio2021, xie2016}, or by convolutional layers acting as a feature extractor from 2D or 3D input data \cite{guo2017}. Similarly to \cite{guo2017}, we adopted a Convolutional AutoEncoder (CAE) architecture to address the reconstruction of the histological patches as a pretext task. However, our CAE differs from the current literature in a specific aspect of the network: the bottleneck. Unlike Guo et al. \cite{guo2017}, who combined flatten operations with fully-connected layers at the middle of the CAE, we introduced a convolutional attention module through a residual connection to enhance the latent space for the downstream clustering task. As observed in Figure \ref{fig_CAE}, we utilized an encoder composed of three stacked convolutional layers with a $3\times3$ receptive field (blue boxes). At the bottleneck, we defined an attention block composed of a tailored autoencoder which allows refining the embedded features in the spatial dimension. Specifically, the proposed module combines $1\times1$ convolutions (green boxes) with a sigmoid function (purple layer) intended to recalibrate the inputs. The inclusion of an identity shortcut forces the network to stabilize the feature space by propagating larger gradients to previous layers via skip connections. An additional $1\times1$ convolutional layer was included to modify the filter channel without affecting the dimension of the feature maps. In the decoder stage, we applied regularization operations between the transpose convolutional layers (yellow-contour boxes) throughout Batch Normalization to avoid the internal covariate shift \cite{ioffe2015}. A remarkable factor is that no pooling or upsampling layers were used to adapt the dimensions of the feature maps after each convolutional step. Instead, we worked with a $stride>1$ both in the encoder and decoder structures to provide a network with a higher capability of transformation by learning spatial subsampling.

As observed in Figure \ref{fig_CAE}, given an input set of patches $X=\{x_1, x_2, ..., x_i, ..., x_N\}$, with $N$ the number of samples per batch, the encoder network maps each input $x_i\in \mathbb{R}^{M\times M\times 3}$ into an embedded feature space $z_i=f_\phi(x_i)$ resulting from the attention module. At the end of the autoencoder network, the decoder function was trained to provide a reconstruction map $r_i=g_\theta(z_i)$ trying to minimize the mean squared error (MSE) between the input $x_i$ and the output $r_i$, according to Equation \ref{eq_mse}. Note that the histological patches were resized from $M_0=512$ to $M=128$ to alleviate the GPU constraints during the training of the model. 

\begin{equation}
    L_r = \frac{1}{N}{\sum_{i=1}^N{||x_i-g_\theta(f_\phi(x_i))||^2}}
\label{eq_mse}
\end{equation}

\textbf{Learning details for the CAE pre-training.}

Let $\mathcal{X}=\{X_1, ..., X_b, ..., X_B\}$ be the training set composed of 2995 histological patches, the proposed CAE was trained during $\epsilon=200$ epochs by applying a learning rate of 0.5 on $B=94$ batches, being $X_b\subset \mathcal{X}$ a single batch composed of $N=32$ samples. Adadelta optimizer \cite{zeiler2012} was used to update the reconstruction weights trying to minimize the MSE loss function $L_r$ after each epoch $e$, as detailed in Algorithm \ref{alg_CAE}.

\begin{algorithm}
\caption{CAE training.}
\label{alg_CAE}
\BlankLine
\KwData{Unlabelled training data set $\mathcal{X}=\{X_1, ..., X_b, ..., X_B\}$}
\textbf{Results:} Trained convolutional autoencoder (CAE) parameters $\phi$ and $\theta$.
\BlankLine
$\phi, \theta \leftarrow $ random\;
\For{$ e \leftarrow 1$ \KwTo $\epsilon$}{
    \For{$ b \leftarrow 1$ \KwTo $B$}{
        $X \leftarrow X_b\subset\mathcal{X}$\;
        \For{$ i \leftarrow 1$ \KwTo $N$}{
            $r_i \leftarrow g_\theta(f_\phi(x_i))$ \;
        }
    }
    $\mathcal{L}_r \leftarrow \frac{1}{N}{\sum_{i=1}^N{||x_i-r_i||^2}}$\;
    Update $\phi, \theta$ using $\nabla_{\phi, \theta}\mathcal{L}_r$
}
\end{algorithm}

\subsection{DCEAC training} \label{subsec:3_2_DCEAC_training}

In the pioneer deep-clustering work \cite{xie2016}, the authors proposed a Deep Embedded Clustering (DEC) algorithm in which the decoder structure was discarded during the second stage corresponding to the clustering training. However, Guo et al. \cite{guo2017} demonstrated that fine-tuning just the encoder network could distort the feature space and hurt the classification performance. Instead, they kept the autoencoder untouched under the statement that AE architectures can avoid embedding distortion by preserving local information of data \cite{peng2016}. For that reason, we also propose a simultaneous learning process for both reconstruction and clustering branches to avoid the corruption of the feature space, similarly to \cite{guo2017}.

\begin{figure*}[h]
\begin{center}
    \includegraphics[width=16cm]{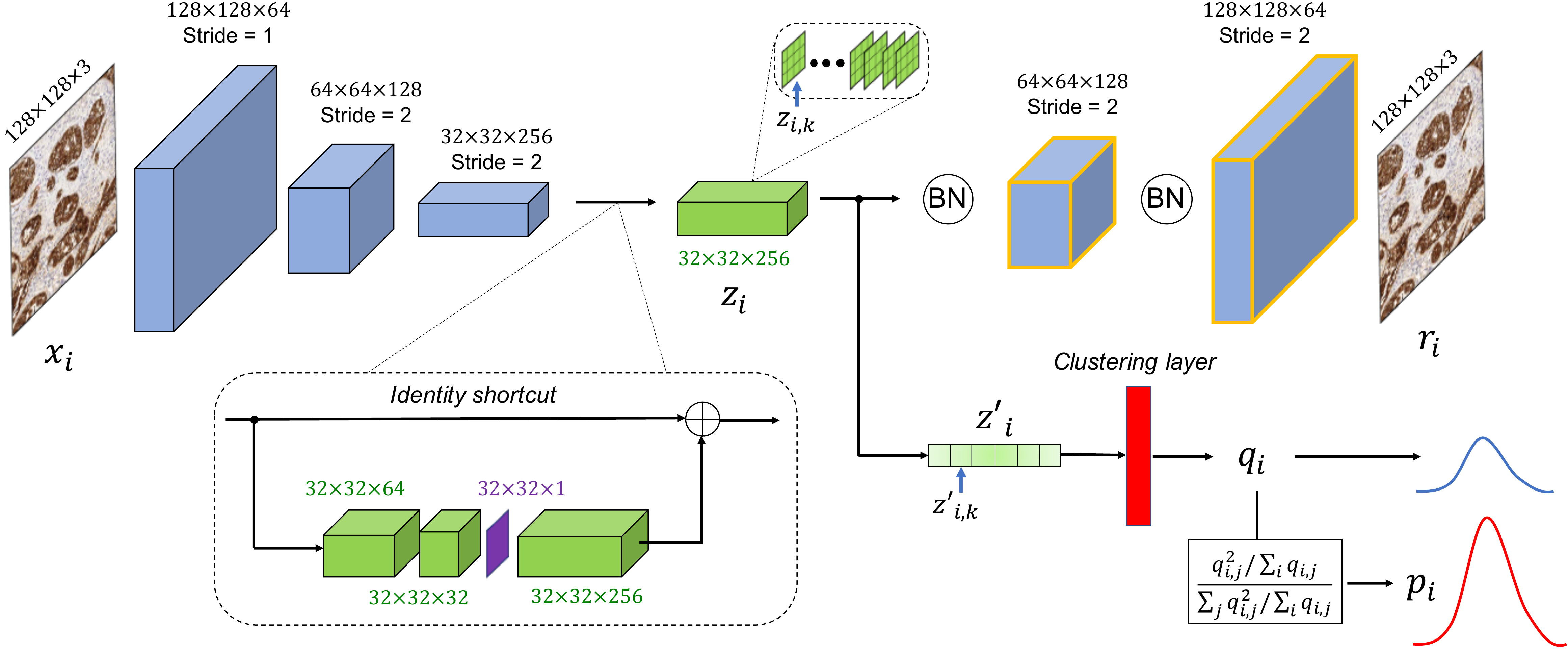}
\end{center}
\caption{Architecture of the proposed Deep Convolutional Embedded Attention Clustering (DCEAC). The model is trained in an end-to-end manner by minimizing both reconstruction and clustering loss functions. The reconstruction pretext task stabilizes the feature space $z_i$ avoiding the embedding distortion, whereas the clustering term predicts the soft-class assignments $q_i$.}
\label{fig_DCEAC}
\end{figure*}

Once the CAE was pre-trained in a first stage (Algorithm \ref{alg_CAE}), we incorporated a clustering branch at the output of the CAE bottleneck giving rise to a Deep Convolutional Embedded Attention Clustering (DCEAC) able to provide a soft label with class dependency. From the embedded representations $z_i = \{z_{i,1}, ..., z_{i,k}, ..., z_{i,C}\}$, being $C=256$ the number of feature maps $z_{i,k}\in \mathbb{R}^{H\times W}$, we performed a spatial squeeze to obtain a feature vector $z'_{i} \in \mathbb{R}^C$ which leads to a better label assignment. As depicted in Figure \ref{fig_DCEAC}, a Global Average Pooling (GAP) layer (fadded green) was used as the projection function to reduce the feature maps $z_{i,k}\in \mathbb{R}^{H\times W}$, with $H=W=32$, into the feature vector $z'_{i,k} \in \mathbb{R}^{1\times 1}$ (see Equation \ref{eq_GAP}).

\begin{equation}
    z'_{i,k} = \frac{1}{H\times W} \sum_{h=1}^{H} \sum_{w=1}^{W} z_{i,k}(h,w)
\label{eq_GAP}
\end{equation}

After the GAP operation, a clustering layer (red box in Figure \ref{fig_DCEAC}) was included to map each embedded representation $z'_i$ into a soft label $q_{i,j}$, which represents the probability of $z'_i$ of belonging to the cluster $j$. According to Equation \ref{eq_q_ij}, $q_{i,j}$ was calculated via Student's t-distribution \cite{van2008} by keeping the cluster centers $\{\mu_j\}_1^K$ as trainable parameters.

\begin{equation}
    q_{i,j} = \frac{(1+||z'_i-\mu_j||^2)^{-1}}{\sum_j(1+||z'_i-\mu_j||^2)^{-1}}
\label{eq_q_ij}
\end{equation}

Note that the cluster centres were initialized by running \textit{k-means} technique on the embedded features $z'_i$, as detailed in Algorithm \ref{alg_DCEAC}. From here, a normal target distribution $p_{i,j}$ (defined in Equation \ref{eq_p_ij}) was used as a ground truth during the training of the models.

\begin{equation}
    p_{i,j} = \frac{q_{i,j}^2/\sum_iq_{i,j}}{\sum_jq_{i,j}^2/\sum_iq_{i,j}}
\label{eq_p_ij}
\end{equation}

The learning framework for the proposed DCEAC (Algorithm \ref{alg_DCEAC}) was conducted by minimizing a custom loss function (Equation \ref{eq_custom_loss}), where $\mathcal{L}_r$ and $\mathcal{L}_c$ are the reconstruction and clustering losses, respectively. $\gamma>0$ is a temperature parameter used to prevent the distortion of the feature space since $\gamma=0$ would be equivalent to train just the convolutional autoencoder (CAE) architecture, as detailed in Section \ref{subsec:3_1_CAE_pretraining}. 

\begin{equation}
    \mathcal{L}=\mathcal{L}_r + \gamma\mathcal{L}_c
\label{eq_custom_loss}
\end{equation}

Specifically, the clustering loss was defined as Kullback-Leibler divergence ($KL=(P||Q)$) according to Equation \ref{eq_KLD}, whereas the mean squared error (mse) was used as a reconstruction loss function.

\begin{equation}
    \mathcal{L}_c = \sum_i\sum_jp_{i_j}log\frac{p_{i,j}}{q_{i,j}}
\label{eq_KLD}
\end{equation}

As mentioned above, autoencoders are in charge of preserving the local structure of data, so the clustering term must provide just a slight contribution to updating the weights in order to avoid the corruption of the latent space. For that reason, we empirically set $\gamma=0.3$ for all the experiments of the training process detailed in Algorithm \ref{alg_DCEAC}.

\begin{algorithm}[h!]
\caption{DCEAC training.}
\label{alg_DCEAC}
\BlankLine
\KwData{Unlabelled training data set $\mathcal{X}=\{X_1, ..., X_b, ..., X_B\}$}
\textbf{Results:} Cluster assignment $\hat{y}_i$ for each histological sample $x_i$.

\BlankLine
\textbf{Step 1: Cluster centers intialization}\\
$\phi, \theta \leftarrow $ pre-trained CAE parameters\;
$\mathcal{Z} \leftarrow f_\phi(\mathcal{X})$\;
$\{\mu_j\}_{j=1}^K \leftarrow kmeans(\mathcal{Z})$\;

\BlankLine
\textbf{Step 2: DCEAC training}\\
\For{$ e \leftarrow 1$ \KwTo $\epsilon$}{
    \For{$ b \leftarrow 1$ \KwTo $B$}{
        $X \leftarrow X_b\subset\mathcal{X}$\;
        \For{$ i \leftarrow 1$ \KwTo $N$}{
            $z_i \leftarrow f_\phi(x_i)$\;
            $r_i \leftarrow g_\theta(z_i)$\;
            $z'_i \leftarrow GAP(z_i)$\;
            $q_{i,j} \leftarrow \frac{(1+||z'_i-\mu_j||^2)^{-1}}{\sum_j(1+||z'_i-\mu_j||^2)^{-1}}$\;
            $p_{i,j} \leftarrow \frac{q_{i,j}^2/\sum_iq_{i,j}}{\sum_jq_{i,j}^2/\sum_iq_{i,j}}$\;
        }
    }
    $\mathcal{L}_r \leftarrow \frac{1}{N}{\sum_{i=1}^N{||x_i-r_i||^2}}$\;
    $\mathcal{L}_c \leftarrow \sum_i\sum_jp_{i_j}log\frac{p_{i,j}}{q_{i,j}}$\;
    $\mathcal{L} \leftarrow \mathcal{L}_r + \gamma\mathcal{L}_c$\;
    Update $\phi, \theta, \mu_j$ using $\nabla_{\phi, \theta, \mu_j}\mathcal{L}$\;
}

\BlankLine
\textbf{Step 3: Label prediction}\\
\For{$ b \leftarrow 1$ \KwTo $B$}{
    $X \leftarrow X_b\subset\mathcal{X}$\;
    \For{$ i \leftarrow 1$ \KwTo $N$}{
        $z'_i \leftarrow GAP(f_\phi(x_i))$\;
        $q_{i,j} \leftarrow \frac{(1+||z'_i-\mu_j||^2)^{-1}}{\sum_j(1+||z'_i-\mu_j||^2)^{-1}}$\;
        $\hat{y}_i \leftarrow argmax_j(q_{i,j})$\;
    }
}
\end{algorithm}

\newpage
\textbf{Learning details for DCEAC training.}

As in the previous CAE pre-training, given an input batch $X_b$ of $N=32$ samples, we made use of Adadelta optimizer with a learning rate of 0.5 to minimize the custom loss function $\mathcal{L}$. Concerning the software and hardware aspects, all models were developed using Tensorflow 2.3.1 on Python 3.6. The experiments were performed on a machine with Intel(R) Core(TM) i7-9700 CPU @3.00GHz processor and 16 GB of RAM. For deep-learning algorithms, a single NVIDIA A100 Tensor Core having cuDNN 7.5 and CUDA Toolkit 10.1 was used.

\section{Experimental results}\label{sec:03_Results}

\subsection{State of the art} \label{subsec:soa_adaptation}

In this section, we show a comparison performance between the proposed DCEAC model and the most relevant deep clustering-based works of the literature. In particular, we adapt the study carried out in \cite{xie2016}, where the authors proposed a two-step learning strategy based on a Deep Embedded Clustering (DEC) model composed of fully connected layers. In the first step, they trained the autoencoder network to extract knowledge from the unlabelled images domain. In the second stage, once the specific-image information was coded, Xie et al. \cite{xie2016} discarded the decoder structure to directly address the clustering phase from the learnt feature space, without considering the reconstruction error. However, posterior works, such as \cite{guo2017}, claimed that convolutional autoencoders (CAEs) are more powerful than fully connected AEs for dealing with images. For that reason, we adapt the previous DEC methodology by including convolution operations instead of fully connected layers. To this end, we follow the methodology exposed in \cite{masci2011}, where stacked CAEs were originally proposed for hierarchical feature extraction. Therefore, in order to conduct a reliable state-of-the-art comparison, we fused both clustering \cite{xie2016} and CAEs architectures \cite{masci2011} to provide a refined DEC model, from now on called \textit{rDEC}.

Otherwise, we also replicated the experiments carried out by Guo et al. \cite{guo2017}, who proposed a hybrid learning for deep clustering with convolutional autoencoders. The main difference with respect to the previous rDEC is that \cite{guo2017} keep untouched the decoder term during the training of the models giving rise to a hybrid framework that combines reconstruction $L_r$ and clustering $L_c$ losses. The idea behind this is that embedded feature space in rDEC could be distorted by only using clustering oriented loss. So, they proposed leveraging the decoder structure to avoid the corruption of the latent space by considering the reconstruction error. Note that one of the main contributions of Guo et al. \cite{guo2017} lied in the proposed bottleneck, since they forced the dimension of the embedded features to be equal to the number of clusters throughout fully connected layers. However, this is not scalable to other classification problems with higher-dimensionality input images or with a small number of clusters. Specifically, they applied the algorithms on the MNIST data set composed of samples $x_i \in \mathcal{R}^{28\times 28\times 1}$ and provide an embedded space $z_i$ with 10 features, according to the $K=10$ number of clusters. However, in our case, we deal with images of $128\times 128\times 3$ pixels, where the high resolution is essential for the classification performance, unlike for MNIST data set. Additionally, we aim to classify the histological samples into $K=3$ classes, so replicating the architecture of \cite{guo2017} is unfeasible since the decoder term would be unable of reconstructing the images just from three feature values. For that reason, to boost a compelling comparison with \cite{guo2017}, we maintain the same architectures and training details proposed in this work, but removing the convolutional attention module for being one of the main own contributions. Hereinafter, we refer to this approach as \textit{rDCEC}.

\subsection{Quantitative results} \label{subsec:quantitative_results}

In this section, we report the unsupervised classification performance achieved by the aforementioned rDEC \cite{xie2016} and rDCEC \cite{guo2017} algorithms in comparison with our proposed DCEAC model. Also, a conventional method based on running \textit{k-means} algorithm on the feature space was considered to know how the gap in performance is between the proposed model and traditional techniques. This conventional approach will be termed as \textit{AE+kmeans}. As observed in Tables \ref{results_per_class} and \ref{average_results}, the comparison is handled by means of different figures of merit, such as sensitivity (SN), specificity (SP), F-score (FS), accuracy (ACC) and area under the ROC curve (AUC). Particularly, Table~\ref{results_per_class} shows results per class to evidence how well the four algorithms classify the 2995 histological patches with non-tumour (NT), mild (M) and infiltrative (I) patterns. Besides, Table~\ref{average_results} reports the classification results in terms of micro and macro-average. Both metrics provide information about the overall average performance of the classification models, but micro-average takes into account the unbalancing between classes, which enables a more faithful perspective of the models' behaviour than macro-average. 

\begin{figure*}[h!]
\begin{center}
    \includegraphics[width=17cm]{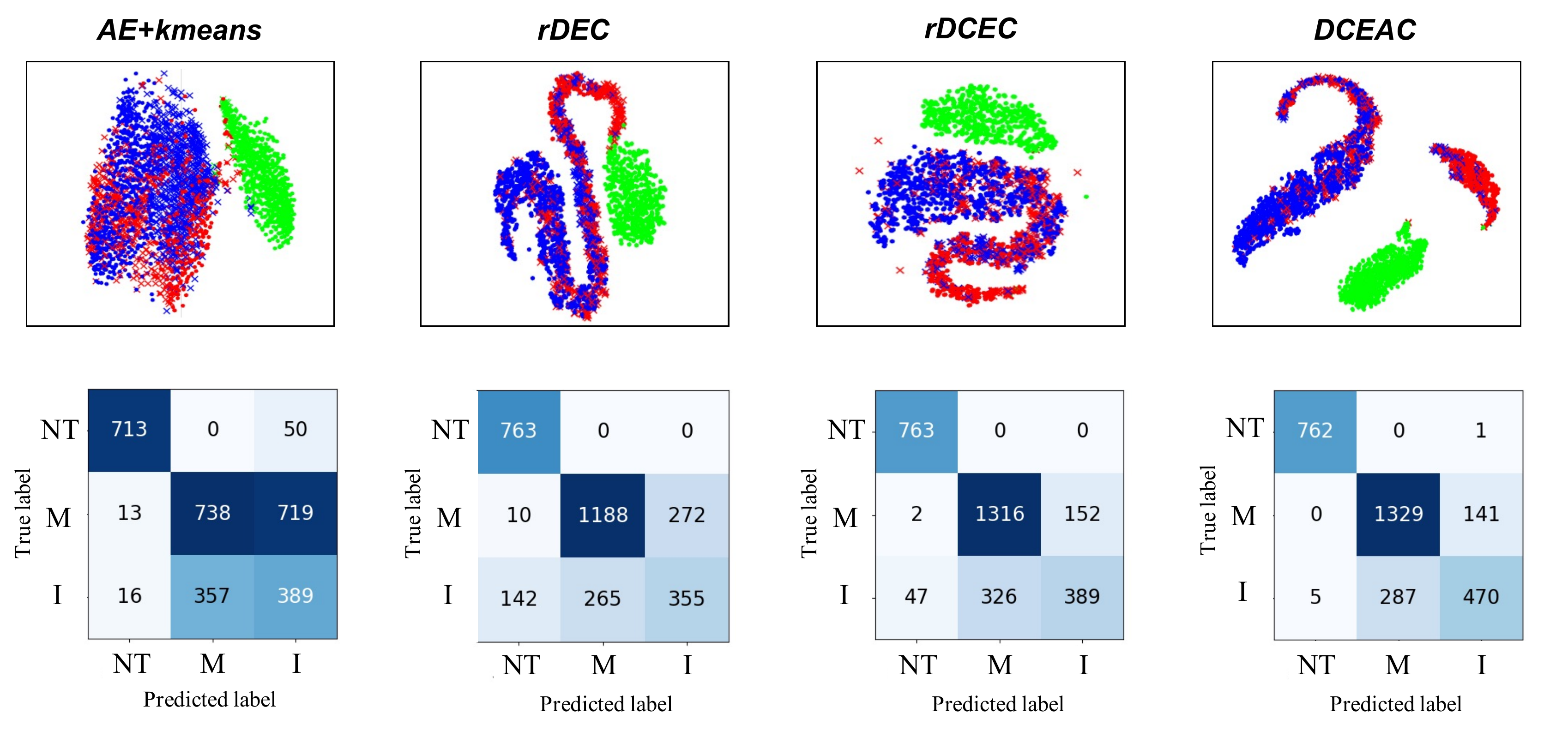}
\end{center}
\caption{Representation of the latent space and confusion matrix derived from the clustering classification reached by each method.}
\label{fig_visual_results}
\end{figure*}

To enhance the comparison between the learning approaches, we represent in Figure \ref{fig_visual_results} the latent space arranged by each model with its respective confusion matrix. Note that, while the confusion matrix gives information about the classification capability of each model, the representation of the embedded space contributes to a more comprehensive clustering scenario for the bladder cancer grading. In this way, the T-distributed Stochastic Neighbor Embedding (TSNE) tool was used to illustrate, in a 2D map, the well and miss-classified embedded features denoted by spots and crosses, respectively. Green, blue and red colours make reference to the non-tumour, mild and infiltrative patterns.

\begin{table*}[h!]
\caption{Unsupervised classification results per class.}
\label{results_per_class}
\centering
\renewcommand{\arraystretch}{1.1}
\resizebox{17cm}{!}{
\begin{tabular}{ccccc|cccc|cccc}
\hline
                                  & \multicolumn{4}{c|}{\textbf{NON-TUMOUR}}                                                     & \multicolumn{4}{c|}{\textbf{MILD}}                                                      & \multicolumn{4}{c}{\textbf{INFILTRATIVE}}                               \\ \hline
                                  & \textit{AE+kmeans} & \textit{rDEC} & \textit{rDCEC} & \multicolumn{1}{c|}{\textit{DCEAC}} & \textit{AE+kmeans} & \textit{rDEC}   & \textit{rDCEC} & \multicolumn{1}{c|}{\textit{DCEAC}} & \textit{AE+kmeans} & \textit{rDEC} & \textit{rDCEC} & \textit{DCEAC}  \\ \hline
\multicolumn{1}{c|}{\textbf{SN}}  & 0.9345             & \textbf{1}    & \textbf{1}     & 0.9987                              & 0.5020             & 0.8082          & 0.8952         & \textbf{0.9041}                     & 0.5105             & 0.4659        & 0.5105         & \textbf{0.6168} \\
\multicolumn{1}{c|}{\textbf{SP}}  & 0.9870             & 0.9319        & 0.9780         & \textbf{0.9978}                     & 0.7659             & \textbf{0.8262} & 0.7862         & 0.8118                              & 0.6556             & 0.8782        & 0.9319         & \textbf{0.9364} \\
\multicolumn{1}{c|}{\textbf{FS}}  & 0.9475             & 0.9094        & 0.9689         & \textbf{0.9961}                     & 0.5754             & 0.8129          & 0.8458         & \textbf{0.8613}                     & 0.4052             & 0.5112        & 0.5971         & \textbf{0.6841} \\
\multicolumn{1}{c|}{\textbf{ACC}} & 0.9736             & 0.9492        & 0.9836         & \textbf{0.9980}                     & 0.6364             & 0.8174          & 0.8397         & \textbf{0.8571}                     & 0.6187             & 0.7733        & 0.8247         & \textbf{0.8551} \\ \hline
\end{tabular}}
\end{table*}

\begin{table*}[h!]
\caption{Unsupervised classification results in terms of micro and macro-average.}
\label{average_results}
\centering
\renewcommand{\arraystretch}{1.1}
\resizebox{13cm}{!}{
\begin{tabular}{ccccc|cccc}
\hline
\textbf{}                         & \multicolumn{4}{c|}{\textbf{MICRO-AVERAGE}}                           & \multicolumn{4}{c}{\textbf{MACRO-AVERAGE}}                           \\ \hline
\textbf{}                         & \textit{AE+kmeans} & \textit{rDEC} & \textit{rDCEC} & \textit{DCEAC}  & \textit{AE+kmeans} & \textit{rDEC} & \textit{rDCEC} & \textit{DCEAC}  \\ \hline
\multicolumn{1}{c|}{\textbf{SN}}  & 0.6144             & 0.7699        & 0.8240         & \textbf{0.8551} & 0.6490             & 0.7580        & 0.8019         & \textbf{0.8399} \\
\multicolumn{1}{c|}{\textbf{SP}}  & 0.8072             & 0.8850        & 0.9120         & \textbf{0.9275} & 0.8028             & 0.8788        & 0.8987         & \textbf{0.9153} \\
\multicolumn{1}{c|}{\textbf{FS}}  & 0.6144             & 0.7699        & 0.8240         & \textbf{0.8551} & 0.6427             & 0.7445        & 0.8039         & \textbf{0.8472} \\
\multicolumn{1}{c|}{\textbf{ACC}} & 0.7429             & 0.8466        & 0.8827         & \textbf{0.9034} & 0.7429             & 0.8466        & 0.8827         & \textbf{0.9034} \\
\multicolumn{1}{c|}{\textbf{AUC}} & 0.7259             & 0.8184        & 0.8503         & \textbf{0.8776} & 0.7259             & 0.8184        & 0.8503         & \textbf{0.8776} \\ \hline
\end{tabular}}
\end{table*}

\subsection{Qualitative results} \label{subsec:qualitative_results}

In an attempt to incorporate an interpretative perspective for the reported quantitative results, we computed the class activation maps (CAMs), which allows highlighting the regions in which the model pays attention to predict the class of each sample. This fact usually can help to find hidden patterns associated with a specific class or to determine if the label prediction is based on the same patterns as clinicians. In this way, the reported heatmaps lead to a better understanding of the embedded feature space by pointing out decisive areas of the histological patches for the cluster assignment. 

\begin{figure*}[h!]
\begin{center} 
    \includegraphics[width=16cm]{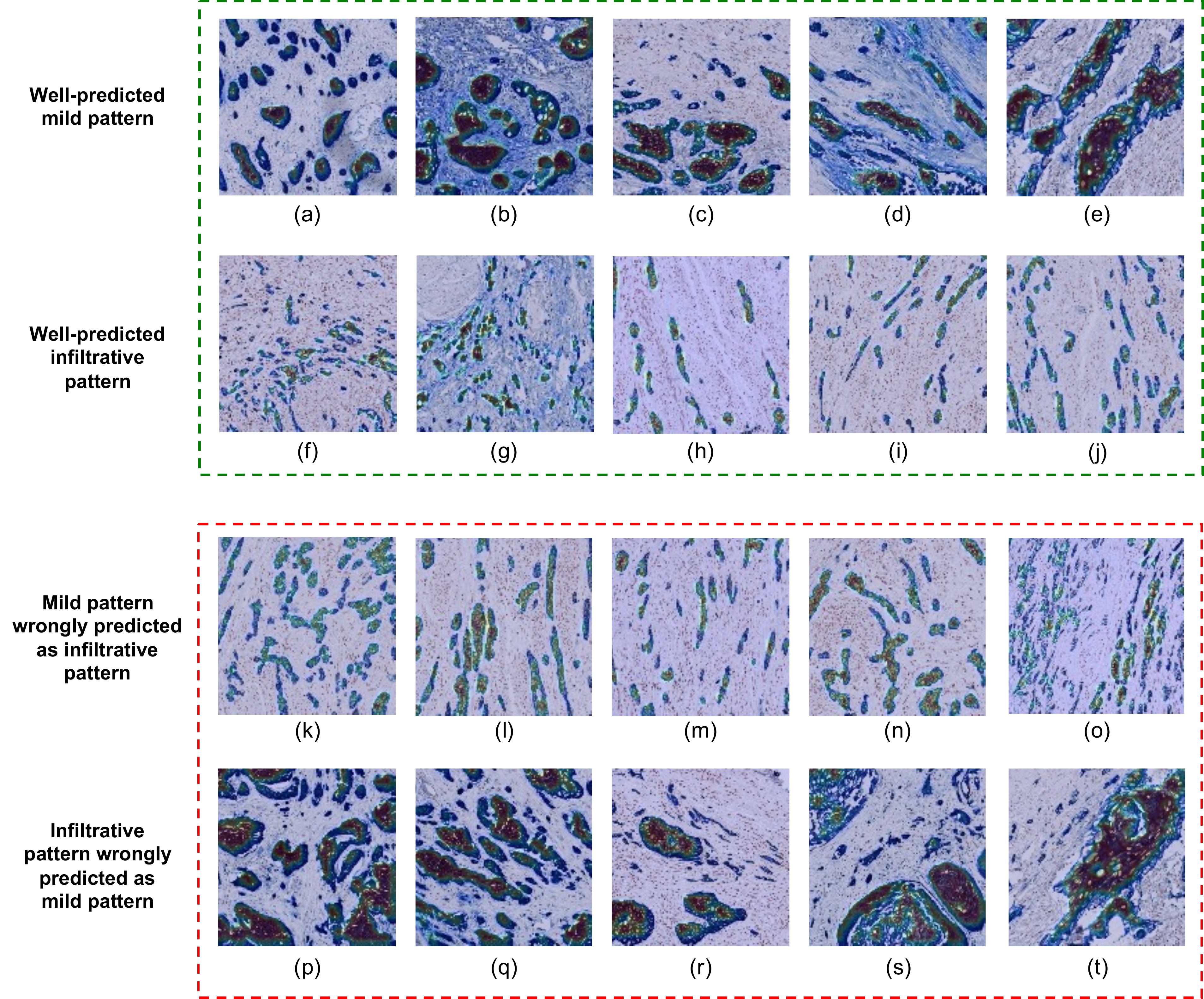}
\end{center}
\caption{Class activation maps highlighting the regions that the proposed DCEAC model considers relevant for the class prediction. The green frame refers to well-predicted images with mild (M) and infiltrative(I) patterns, whereas the red frame corresponds to the miss-classified samples in which the aggressiveness of the disease has been confused.}
\label{fig_CAMs}
\end{figure*}

As deduced from Figure \ref{fig_visual_results}, the biggest challenge of the bladder muscle-invasive cancer (MIBC) grading lies in the distinction of the mild (M) and infiltrative (I) cancerous patterns, as expected. For that reason, in Figure \ref{fig_CAMs}, we report several examples of heatmaps corresponding to miss-classified samples to elucidate the reason why the proposed model wrongs. Also, we show examples of well-predicted CAMs to evidence the relevant structures in which the network pays attention when correctly predicting. Specifically, we show five examples per case to make clear the criteria followed by the proposed model to determine the class. In the green frame of Figure \ref{fig_CAMs}, we illustrate well-classified mild (a-e) and infiltrative (f-j) histological patterns. Additionally, in the red frame, we show bladder cancer samples with a mild pattern miss-classified as tumour budding (k-o), and vice versa (p-t). The findings from the class activation maps will be discussed in Section \ref{sec:05_Discussion}.

\section{Discussion}\label{sec:05_Discussion}

\subsection{About quantitative results}

From Table \ref{results_per_class}, we can observe that all the contrasted models work well for detecting the non-tumour class, including the conventional \textit{kmeans}. However, the proposed DCEAC model reaches the higher performance for all the metrics, except for sensitivity since the model miss-classifies one non-tumour sample, as reported in the DCEAC's confusion matrix of Figure \ref{fig_visual_results}. Regarding the mild class, the proposed model also provides the best behaviour. Only rDEC surpass a $1\%$ specificity, but at the cost of compromising a $10\%$ the sensitivity concerning the proposed DCEAC. As observed in Table \ref{results_per_class}, the outperforming of our model is even more remarkable when discerning the infiltrative pattern. DCEAC shows the best results for all the metrics, especially for sensitivity and F-score exceeding more than a $10\%$ to the rest of the approaches. 

Table \ref{average_results} reports the overall performance of the models, in terms of micro and macro-average. As mentioned above, the micro-average results take into account the unbalancing between classes, which is an important aspect in this study since the samples with mild pattern appears oversampled. Nevertheless, the proposed DCEAC model consistently outperforms the rest of the clustering methods by $2-3\%$ across the different both micro and macro-average metrics, as appreciated in Table \ref{average_results}. As the final remark concerning the quantitative results, it should be highlighted that the expert's decision coincides with the proposed artificial intelligence system in the $90,34\%$ of the cases, according to the average accuracy. 

A reinforcement of the quantitative results is reported in Figure \ref{fig_visual_results}. From the confusion matrices, it is clearly appreciated by the colour range that all the models tend to confuse mild and infiltrative cancerous patterns. The \textit{kmeans} algorithm presents a very low capability of discerning between carcinogenic samples since most of the images are predicted as a mild pattern because of the oversampling of that class. This changes when deep-clustering algorithms are outlined. In particular, the rDEC model improves the classification of carcinogenic images but compromises a lot the non-tumour class by miss-classifying samples with an infiltrative pattern. In this line, the rDCEC model improves the results by notably decreasing the tumour budding samples wrongly predicted as non-tumour cases. In addition, rDCEC also increases the number of true positives for tumour samples. However, this model presents a significant lack when predicting the infiltrative patterns, since an important number of them are wrongly labelled as mild. Unequivocally, the proposed DCEAC model provides the best classification results. The samples with tumour budding miss-classified as non-tumour cases decreases almost to a minimum, unlike in previous cases. Moreover, the number of true positives for mild and infiltrative patterns reaches the highest values, while reducing the false positives and negatives.

The representation of the embedded feature space offers a visual perspective for the quantitative results. We can observe that \textit{kmeans} algorithm is able to roughly discern between non-tumour and carcinogenic histological samples. However, the point cloud is very diffuse to separate mild and infiltrative classes. Contrarily, the rDEC model shows a better distribution of the embedded data, although the features relative to each class still remain close together in the latent space. This improves in the case of the rDCEC model, where independent clusters begin to be appreciated. The non-tumour features (denoted by the green colour) are unmarked in the representation space and the embedded tumour samples begin to scatter towards different cluster classes. Indisputably, the proposed DCEAC model provides the best embedding representation since the features are distributed along the latent space forming independent clusters according to a specific class. This fact further strengthens our confidence in the ability of the proposed model to discern between non-tumour, mild and infiltrative histological patterns. 

From the previous in-depth analysis of the quantitative and interpretative results, we can clear up several findings. The first of them is that the use of deep-learning techniques improves the classification performance regarding conventional clustering approaches. As expected, all the deep clustering-based methods, i.e. rDEC, DCEC and DCEAC, outperform the baseline based on the \textit{kmeans} algorithm. This is because these models enable a deeper learning stage in which the embedded features are adjusted to a target distribution, unlike the \textit{kmeans} algorithm which modifies the clusters iteratively without updating the feature learning. Additionally, we can observe that models with both reconstruction and clustering branches integrated into a unified framework provide better results than the rDEC model, which addresses the learning process in two independent stages. The reason behind the outperforming of rDCEC and DCEAC with respect to rDEC lies in the preservation of the local structure of the embedded data. Since rDCEC and DCEAC models have a connected output between the clustering and reconstruction stages, the clustering term can transfer class information to the reconstruction term, which is in charge of updating the weights of the encoder network. In this way, the embedded features can be optimized by incorporating the class prediction without distorting the latent space thanks to the decoder structure. Finally, the proposed DCEAC model showcases substantial performance improvements regarding the rest of the approaches. This is due to the inclusion of the convolutional attention block, which allows refining the latent space to provide more suitable features for the clustering phase.

\subsection{About qualitative results}

As appreciated in the class activation maps (CAMs) reported in Figure \ref{fig_CAMs}, the proposed DCEAC model focuses on tumour cell nests (Figure \ref{fig_CAMs} (a-c)) and tumour interconnected bands (Figure \ref{fig_CAMs} (d-e)) when predicts samples with a mild pattern. This fact implies that the proposed network has learnt by itself to associate nodular and trabecular structures with a mild pattern of the disease. Additionally, DCEAC model recognises small sets of isolated buds (Figure \ref{fig_CAMs} (f-g)) or tumour cell strands (Figure \ref{fig_CAMs} (h-j)) as characteristic structures of the infiltrative pattern, a.k.a. tumour budding. These findings are evidenced in the green frame of the heat maps corresponding to well-predicted samples. 

In the case of the wrong predictions (red frame in Figure \ref{fig_CAMs}), we can observe that the proposed network maintains consistency when determining the class of each sample. The histological patches of Figure \ref{fig_CAMs} (k-o) show an appearance more similar to the infiltrative patterns, so the network highlights small strands reminiscent of tumour budding structures. However, the true label assigned by the expert for these samples was a mild pattern. At this point, the qualitative results become interesting because the model' suggestions can lead the pathologists to reconsider its diagnosis in some doubtful cases, as a second opinion. Besides, the human eye is susceptible to fatigue, so the proposed system could help in cases where some patterns have gone unnoticed, in order to avoid a biased diagnosis.

Otherwise, in the cases of the Figure \ref{fig_CAMs} (p-t), samples with an infiltrative pattern are wrongly predicted by the model as mild cases. In these histological patches, the proposed network focuses on bigger structures related to nodular or trabecular patterns, but it ignores the small isolated tumour cells that lead to higher severity of bladder cancer. From here, we can deduce that, although the final prediction could be wrong, the pattern recognition accomplished by the model maintains consistency. Notice that the model wrongs because patterns belonging to a different class coexist in the same histological patch, so in future research lines, we will face this long-standing problem. 

In summary, the proposed DCEAC model demonstrates, via class activation maps, a high-confident prediction since it is able to focus on the same patterns as clinicians, without having previous information from them. As mentioned above, the opinion from the expert and the proposed model will match in most of the cases (in concrete, $90,34\%$ of the time). Thus, the artificial intelligence system could help as a computer-aided system for reviewing processes, which would give rise to an improvement of the diagnosis quality without involving other experts. Also, the proposed system could help inexperienced pathologists by suggesting areas of interest with a convincing label. 

\section{Conclusion}\label{sec:06_Conclusion}

In this paper, we have proposed a novel self-learning framework based on deep-clustering techniques to grade the severity of bladder cancer through histological samples. Immunohistochemistry staining methods were applied on the images to enhance the non-tumour, mild and infiltrative patterns, according to the literature. We resorted to a hybrid model by combining a clustering branch along the reconstruction term used as a pretext task to preserve the local structure of the features. To stand out from the state of the art, we introduced a convolutional attention block that allows refining the feature space to lead to a better-unsupervised classification. The proposed Deep Convolutional Embedded Attention Clustering (DCEAC) has demonstrated outperforming previous clustering-based methods, achieving an average accuracy of $0.9034$ to grade the aggressiveness of the muscle-invasive bladder cancer (MIBC). Additionally, the reported class activation maps (CAMs) show that the proposed system is able to learn by itself the same structures as clinicians to associate the patterns with the correct severity degree of the disease, without incurring prior annotation steps. In this line, our fully unsupervised perspective bridges the gap with respect to other supervised algorithms, since the proposed system does not require expert involvement to be trained.  

In future research lines, we will work on improving the accuracy of tumour samples when structures of different growth patterns appear on the same image. In addition, we will use more powerful hardware systems to process entire high-resolution Whole Slide Images to provide a diagnosis per biopsy, instead of per patch.



\section*{Funding}

This work has been partially funded by SICAP project (DPI2016-77869-C2-1-R) and GVA through project PROMETEO/2019/109. The work of Gabriel Garc\'{i}a has been supported by the State Research Spanish Agency PTA2017-14610-I. The equipment used for this research has been funded by the European Union within the operating Program ERDF of the Valencian Community 2014-2020 with the grant number IDIFEDER/2020/030.







\bibliographystyle{elsarticle-num}
\bibliography{refs.bib}







\end{document}